\documentstyle[aps,prl,twocolumn,epsfig]{revtex}

\begin{document}

\twocolumn[\hsize\textwidth\columnwidth\hsize\csname@twocolumnfalse\endcsname

\title{Controlling the cold collision shift in high precision atomic interferometry}
\author{F. Pereira Dos Santos, H. Marion, S. Bize$^{a}$, Y. Sortais, and A. Clairon\\
BNM-SYRTE, Observatoire de Paris, 61 Avenue de l'Observatoire, 75014 Paris, France \\
C. Salomon\\
Laboratoire Kastler Brossel, ENS, 24 rue Lhomond, 75005 Paris,
France}

\date{\today}
\maketitle

\begin{abstract}

We present here a new method based on a transfer of population by
adiabatic passage that allows to prepare cold atomic samples with
a well defined ratio of atomic density and atom number. This
method is used to perform a measurement of the cold collision
frequency shift in a laser cooled cesium clock at the percent
level, which makes the evaluation of the cesium fountains accuracy
at the $10^{-16}$ level realistic. With an improved set-up, the
adiabatic passage would allow measurements of atom
number-dependent phase shifts at the $10^{-3}$ level in high
precision experiments.

\


\

\end{abstract}]

Collisions play an important role in high precision atomic
interferometry \cite{Chin01,Bijlsma94}. In most experiments, a
precise control of the atomic density is hard to achieve, which
sets a limit to how accurately systematic effects due to
collisions can be corrected for. This is particularly true for
clocks using laser cooled Cs atoms \cite{Gibble93,Ghezali96}. The
accuracy of the BNM SYRTE cesium fountains reaches now
$\sim1\times10^{-15}$. It is currently limited by a 10 to 20 \%
systematic and statistical uncertainty on the determination of the
cold collision shift \cite{Lemonde01,Sortais01,Bize01StAndrews}.
To reach such an accuracy, one has to operate with a moderate
number of detected atoms, typically $10^{5}$, which sets a
standard quantum limit to the frequency stability
\cite{Santarelli99} of about $10^{-13}\tau^{-1/2}$, where $\tau$
is the averaging time in seconds. However, when using a high
number of atoms ($10^6-10^7$), a stability approaching
$3\times10^{-14}\tau^{-1/2}$ has already been demonstrated
\cite{Bize01StAndrews,Santarelli99}, which would make the
evaluation at the $10^{-16}$ level practicable. Under these
conditions, the cold collision frequency shift is very large
($\sim10^{-14}$). To actually reach such an accuracy and to take
full advantage of this capability, this shift has to be determined
more accurately than presently achieved.

In this letter, we present a method using adiabatic passage (AP)
\cite{Messiah,Loy74} that allows to prepare atomic samples with
well defined density ratios. This enables the determination of the
collisional frequency shift at the percent level, or better.

The measurement of the cold collision shift is based on a
differential method \cite{Sortais00}. One alternates sequences of
measurements with two different, high and low effective atomic
densities. One then measures the frequency difference between the
two situations, as well as the difference in the number of
detected atoms. Knowing the effective densities, the clock
frequency can be corrected for the collisional shift by
extrapolation to zero density. Unfortunately, the effective
density cannot be measured directly : in a fountain, one only
measures the number of detected atoms. A full numerical simulation
(which takes into account the whole geometry of the fountain and
the parameters of the atomic cloud) is then necessary to estimate
the effective density, as in \cite{Sortais00}. Nevertheless,
extrapolating the clock frequency to zero density is still
possible if one assumes that the effective density and the number
of detected atoms are proportional. Under these conditions, the
collisional frequency shift $\delta \nu$ is proportional to the
number of detected atoms $N_{det}$, say $\delta \nu=K N_{det}$.
The important point is that the coefficient $K$ should be the same
for the low and high density configurations, otherwise the
extrapolation to zero detected atoms is inaccurate.

Up to now, two methods have been used to change the density, and
hence $N_{det}$. Atoms are initially loaded in an optical
molasses, whose parameters (duration, laser intensity) can be
varied. A better technique consists in keeping the same loading
parameters but changing the power in a selection microwave cavity,
which is used to prepare atoms in the
$\left|F=3,m_F=0\right\rangle$ state. One can select all (resp.
half) of the atoms initially in the $\left|F=4,m_F=0\right\rangle$
state by applying a $\pi$ (resp. $\pi/2$) pulse. However, due to
the microwave field inhomogeneities in the cavity, the pulses
cannot be perfectly $\pi$ and $\pi/2$ pulses for all the atoms.
Both techniques affect the atomic densities, velocity distribution
and collisional energy \cite{Leo01}, and consequently the $K$
coefficients usually differ for the low and high density cases.
Numerical simulations show that the $K$ coefficient may differ by
10\% to 15\% in our set-up, depending on parameters such as
microwave power, velocity and position distribution. Fluctuations
and imperfect determination of those parameters prevent from
performing an accurate evaluation of the $K$ coefficient.

A method immune of these systematic effects prescribes to change
the number of atoms of the sample without changing neither its
velocity distribution, nor its size. This can be realized by an
adiabatic transfer of population, which allows one to prepare two
atomic samples, where both the ratio of the effective densities
and the ratio of the atom numbers are exactly $1/2$. In contrast
to previous methods, this one is insensitive to fluctuations of
experimental parameters such as the size and temperature of the
atomic sample, or the power coupled into the selection cavity.

\begin{figure}[htb]
\begin{center}
\epsfig{file=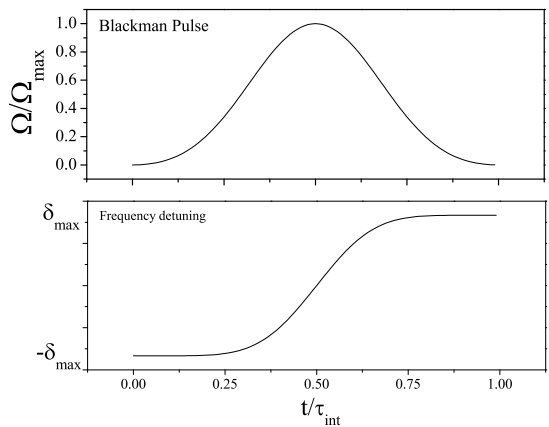,height=6cm}
\end{center}
\caption{\footnotesize Temporal dependence of the Blackman pulse
and the corresponding frequency chirp.} \label{fig:pulses}
\end{figure}

First, an adiabatic passage in the selection cavity is used to
transfer with a 100\% efficiency all the atoms from the
$\left|F=4,m_F=0\right\rangle$ state to the
$\left|F=3,m_F=0\right\rangle$ state \cite{Messiah,Loy74}. This
requires that the microwave field in the cavity is swept across
resonance, and that the Rabi frequency $\Omega/2\pi$ has an
appropriate shape and maximum intensity. We choose to use Blackman
pulses (BP), following \cite{Kuhn97}, which minimizes
off-resonance excitation. In order to fulfill the adiabaticity
condition, the frequency chirp $\delta$ has to be shaped according
to
\begin{equation}
\delta (t) \propto \int {\Omega}^2(t)dt.
\end{equation}
Figure \ref{fig:pulses} shows the evolution of the microwave field
amplitude together with the frequency chirp.

Second, we exploit another striking property of AP. If we stop the
AP sequence when $\delta=0$ (half-Blackman pulse HBP), the atoms
are left in a superposition of the $\left|F=4,m_F=0\right\rangle$
and $\left|F=3,m_F=0\right\rangle$ states, with weights
rigourously equal and independent of the Rabi frequency. After
removal of the $\left|F=4\right\rangle$ atoms with a pushing laser
beam, half of the atoms are in the $\left|F=3,m_F=0\right\rangle$
state, as desired.

\begin{figure}[htb]
\begin{center}
\epsfig{file=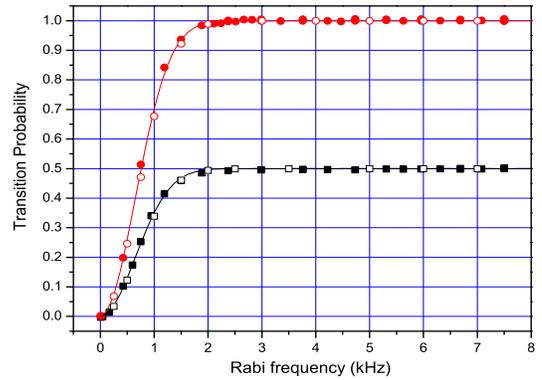,height=5cm,width=7cm}
\end{center}
\caption{\footnotesize Transition probabilities for a BP and a HBP
with $\tau_{int}=4$ ms, $\delta_{max}=5$ kHz as a function of
$\Omega_{max}/2\pi$. The results of the numerical simulations are
displayed as lines (homogeneous Rabi frequency case) and open
symbols (TE011 cavity case), whereas the measurements are full
symbols.} \label{fig:probatranspuis}
\end{figure}

In order to optimize this AP method and to evaluate its
sensitivity to experimental parameters, we first performed a
simple numerical simulation, solving the time-dependent
Schr$\ddot{\rm{o}}$dinger equation for a two level atom in a
homogeneous microwave field. The choice of the pulse parameters
comes from a compromise between the insensitivity of the
transition probabilities to fluctuations and the parasitic
excitation of non-resonant transitions. Figure
\ref{fig:probatranspuis} displays as lines the calculated
transition probabilities as a function of the maximum Rabi
frequency $\Omega_{max}/2\pi$, for BP and HBP. The parameters were
a duration $\tau_{int}=4$ ms, and $\delta_{max}=5$ kHz, which were
constant over the course of the experiment. The simulation shows
that the transition probabilities deviate from 1 and 1/2 by less
than $10^{-3}$ as soon as $\Omega_{max}/2\pi$ is larger than 2.4
kHz. A more realistic simulation has been performed, taking into
account the gaussian spatial distribution of the atomic cloud
(characterized by $\sigma=3.5$ mm) in the selection cavity, its
trajectory in the microwave cavity, as well as the microwave field
distribution of the TE$_{011}$ mode for our cylindrical cavity.
The simulation indicates that the transition probability deviates
from the ideal value by less than $10^{-3}$ for both BP and HBP
for all the atoms contained within $\pm3.5\sigma$ of the vertical
spatial distribution (more than 99.95 \% of the atoms in the
cloud). In this calculation, $\delta=0$ when the center of the
cloud reaches the center of the cavity, which minimizes the
sensitivity of the transfer efficiency to microwave field
inhomogeneities and timing errors. For instance, a delay as large
as 1 ms with respect to the optimal timing induces only a
$7\times10^{-5}$ variation on the transition probability. The only
critical parameter is the accuracy of the frequency at the end of
the chirp $\delta_0$, for HBP. We calculate a linear sensitivity
of the transition probability to $\delta_0$ of
$7\times10^{-5}$/Hz.

We use one of our Cs fountain to demonstrate the AP method and the
resulting ability to control the collisional shift. This clock is
an improved version of the Rb fountain already described elsewhere
\cite{Bize01StAndrews,Bize99}. We use a laser slowed atomic beam
to load about $10^9$ atoms within 800 ms in a lin $\perp$ lin
optical molasses, with 6 laser beams tuned to the red of the
$F=4\rightarrow F'=5$ transition at 852 nm. The atoms are then
launched upwards at $\sim 4.1$ m/s within 2 ms, and cooled down to
an effective temperature of $\sim 1\mu$K. After launch, the atoms
are prepared into the $\left|F=3,m_F=0\right\rangle$ state using a
combination of microwave and laser pulses : they first enter a
selection cavity ($Q\sim$1000) tuned to the
$\left|F=4,m_F=0\right\rangle\rightarrow\left|F=3,m_F=0\right\rangle$
transition, where they experience either BP or HBP pulses. The
atoms left in the $F=4$ state are pushed by a laser beam tuned to
the $F=4\rightarrow F'=5$ transition, 10 cm above the selection
cavity. The amplitude of the pulses are shaped by applying an
adequate voltage sequence (500 steps) to a microwave
voltage-controlled attenuator (60 dB dynamic range), whereas the
frequency chirp is performed with a voltage controlled oscillator.
The Rabi frequency profile agrees with the expected Blackman shape
within a few percent. The frequency chirp, and more specifically
its final frequency, was not controlled as it cannot be easily
checked at the required precision level of 10 Hz for HBP. The
selected atoms then interact with a 9.2 GHz microwave field
synthesized from a high frequency stability quartz oscillator
weakly locked to the output of a H-maser. The two $\pi$/2 Ramsey
interactions are separated by 500 ms. The number of atoms
$N_{F=3}$ and $N_{F=4}$ are finally measured by time of flight
fluorescence, induced by a pair of laser beams located below the
molasses region. From the transition probabilities
$N_{F=4}/(N_{F=3}+N_{F=4})$ measured on both sides of the central
Ramsey fringe, an error signal is computed to lock the microwave
interrogation frequency to the atomic transition using a digital
servo loop.

The transition probabilities are first measured as a function of
the maximum Rabi frequency $\Omega_{max}$, for the Blackman and
half-Blackman pulses. The atoms are launched and selected with the
pushing beam off for this evaluation phase only. To reject the
fluctuations of the initial number of atoms, we measure the ratio
of the atoms transferred into $\left|F=3,m_F=0\right\rangle$ and
the total number of launched atoms, in all magnetic sub-levels. We
then rescale the transfer probability in between 0 and 1 using
only one free parameter : the initial population of the
$\left|F=4,m_F=0\right\rangle$ state. The results are shown in
figure \ref{fig:probatranspuis} and reproduce very well the
numerical simulations.

As the maximum Rabi frequency during the experiment was set to
$7.5$ kHz, the resonance frequencies for transitions between
$m_F\neq0$ states have to be significantly shifted away from the
0-0 transition. A magnetic field of $\sim 180$ mG is applied
during the pulses which keeps the parasitic excitation of magnetic
field sensitive transitions below 0.3 \%. This pulse induces a
quadratic Zeeman shift on the 0-0 transition of about 14 Hz than
must be taken into account to meet the resonance condition
$\delta=0$ for HBP.

\begin{figure}[htb]
\begin{center}
\epsfig{file=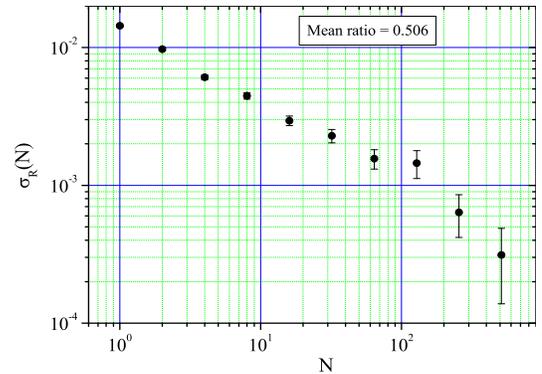,height=5cm,width=7cm}
\end{center}
\caption{\footnotesize Standard deviation of the fluctuation of
the ratio $R(N)$.}\label{fig:ratio}
\end{figure}

For each sequence of the differential measurement, we measure the
mean atom number for the Blackman and half-Blackman pulses, and
compute their ratio $R$. We then calculate $R_N$, the average of
$R$ for N successive sequences. In figure \ref{fig:ratio}, the
standard deviation $\sigma_R(N)$ for various N is plotted. The
stability of $R$ reaches $3\times10^{-4}$ after a one-day
integration. This reflects the insensitivity of the AP to the
experimental parameter fluctuations. The mean value of the ratio
is $R=0.506$, whereas it was expected to be 0.5 at the $10^{-3}$
level. This deviation cannot be explained by a non-linearity of
the detection, which could arise from absorption in the detection
beams. When the absorption in the detection laser beams is changed
by a factor 2, we observe no change in the ratio larger than
$10^{-3}$. We attribute this deviation to the uncertainty in the
final frequency of the sweep. In our present set-up, the sweep is
generated by an oscillator whose accuracy is limited to 50 Hz for
a frequency sweep from -5 to +5 kHz (this difficulty can be solved
by using a dedicated DDS numerical synthesizer). We measure a
linear deviation in the transition probability of
$7.5(3)\times10^{-5}$/Hz in agreement with the predicted value.
This can explain a deviation of the ratio by about
$4\times10^{-3}$. However, it is important to notice that even
when the final frequency is detuned by 50 Hz, the spatial
variation of the transition probability across the atomic sample
is less than $10^{-3}$. All the tests performed here demonstrate
that AP is at least accurate at the 1\% level.

Measurements of the collisional frequency shift are then carried
out using BP and HBP pulses with a large number of atoms
($N_{det}\sim10^7$) in order to amplify the collisional shift.
From a differential measurement, one can extrapolate the frequency
of the clock at zero density with respect to the H-maser. The
relative resolution of the frequency difference is
$2\times10^{-13}\tau^{-1/2}$, limited by the phase noise of the
quartz oscillator used for the fountain interrogation. To check
whether this extrapolation is correct, we measure the corrected
frequency for two different initial temperatures of the atomic
cloud, 1.1 and 2.3 $\mu$K, for which the effective densities,
number of detected atoms and K coefficients are expected to be
different. We switch every 50 cycles between four different
configurations : 1.1 $\mu$K and BP, 1.1 $\mu$K and HBP, 2.3 $\mu$K
and BP, 2.3 $\mu$K and HBP. This rejects long term fluctuations in
the experiment induced by frequency drift of the H-maser used as a
reference, variation of the detection responsivity, and
fluctuations of other systematic effects.

The results are summarized in table \ref{tab:results}. For each
configuration, the measurement of the clock frequency and number
of atoms is averaged over a total time of about 50 hours. One can
then extract the differential collisional shifts with a relative
resolution of $5\times10^{-16}$. The K constants are thus
determined with an uncertainty of about 1 \%. They are found to
differ by about 20 \%. The difference between the corrected
frequencies can then be estimated. The uncertainty on this
measurement is two-fold, a statistical uncertainty, and a
systematic error which reflects the 1 \% uncertainty on the ratio
$R$. We find a difference between the corrected frequencies of
-0.012(7)(5) mHz, which is less than 2 \% of the collisional shift
at high density, and compatible with zero within its error bars.

Table \ref{tab:results2} displays the results obtained using
either the standard selection method (SSM) or AP, for a
temperature of $1.1\mu K$. The $K$ coefficients are found to
differ by about 10\%. Indeed, when using a $\pi/2$ pulse with the
standard selection, the density distribution is transversally
distorted : atoms along the symmetry axis of the cavity are more
efficiently transferred than off-axis atoms. This increases the
effective density for the same number of detected atoms with
respect to AP, giving a larger collision shift at low density. In
fact, $K$ is expected to be lower with SSM than AP, in agreement
with our measurement. Extrapolating the frequency to zero density
when using SSM then leads to an error of about $3\times10^{-15}$
at this density.

It is also important to notice that we measure simultaneously the
collisional frequency shift and the shift due to cavity pulling
\cite{Bize01}, which is proportional to the number of atoms
crossing the Ramsey cavity. Both are correctly evaluated by our
method.

In conclusion, we demonstrate here a powerful method based on an
adiabatic transfer of population to prepare atomic samples with a
well-defined density ratio. An important point is that the cold
collisional shift is measured precisely without any absolute
calibration, nor numerical simulation. This holds even when
parameters of the atomic sample, or even of the atomic detection
are fluctuating. This method can lead to a potential control of
the cold collision shift at the $10^{-3}$ level, or even better.
This capability could be demonstrated by using an ultra stable
cryogenic oscillator \cite{Santarelli99}, allowing a frequency
resolution of $10^{-16}$ per day. Having now at hand a powerful
method to determine the collisional shift with the required
resolution, the evaluation of the Cs fountain accuracy at the
$10^{-16}$ level is reachable. Any other high precision
measurement using cold atoms should also benefit from this method
to evaluate phase shifts induced by atomic interactions.

\

\noindent {\bf Acknowledgments:} The authors wish to thank D.
Calonico for his contribution in previous stage of the experiment,
A. G\'erard and G. Santarelli for technical assistance, and P.
Lemonde for fruitful discussions. This work was supported in part
by BNM and CNRS. BNM-SYRTE and Laboratoire Kastler-Brossel are
Unit\'es Associ\'ees au CNRS, UMR 8630 and 8552.

$^a$ Present address: Time and Frequency Division National
Institute of Standards and Technology 325, Broadway Boulder,
Colorado 80305, USA

\begin{table}[htb]
\caption{Measurement of the cold collisional frequency shift using
Adiabatic Passage. $T$ is the atomic temperature, $R$ the ratio of
the number of detected atoms. The first (resp. second, when
present) error bar indicated in parenthesis reflects the
statistical (resp. systematic) uncertainty.}
\begin{center}
\begin{tabular}{cccc}
$T$ & $\delta \nu$ (mHz) & $R$ & $K$ ($\times10^{-11}$) Hz/at  \\
\hline $1.1 \mu$K &  -0.323(5) & 0.5063(3) & {\bf -8.62(13)} \\
$2.3 \mu$K &  -0.260(5) & 0.5056(3) & {\bf -10.04(20)} \\
\end{tabular}
Difference in corrected frequency : -0.012(7)(5) mHz
\end{center}
\label{tab:results}
\end{table}
\begin{table}[htb]
\caption{Comparison between the Adiabatic Passage technique (AP)
and the Standard Selection Method (SSM). The temperature of the
sample for these measurements was $1.1\mu$K.}
\begin{center}
\begin{tabular}{cccc}
 & $\delta \nu$ (mHz) & $R$ & $K$ ($\times10^{-11}$) Hz/at \\
\hline SSM & -0.234(7)  & 0.540(3) & {\bf -8.08(24)} \\
AP &  -0.275(8) & 0.5054(8) & {\bf -8.97(23)} \\
\end{tabular}
\end{center}
\label{tab:results2}
\end{table}

\end{document}